\documentclass[twocolumn,amsmath,floatfix,amssymb,aps,prl,showpacs,letter,superscriptaddress]{revtex4-1}
\usepackage{graphicx}
\usepackage{color}
\usepackage{graphicx}  
\usepackage{dcolumn}   
\usepackage{bm}        
\usepackage{amssymb}   
\usepackage{amsmath}
\usepackage{gensymb}   
\usepackage{color}
\usepackage{bbm}
\usepackage{hyperref}

\begin{document}

\title{Gapped and gapless short range ordered magnetic states with $(\frac{1}{2},\frac{1}{2},\frac{1}{2})$ wavevectors in the pyrochlore magnet Tb$_{2+x}$Ti$_{2-x}$O$_{7+\delta}$}

\author{E. Kermarrec}
\affiliation{Department of Physics and Astronomy, McMaster University, Hamilton, Ontario, L8S 4M1, Canada}
\author{D. D. Maharaj}
\affiliation{Department of Physics and Astronomy, McMaster University, Hamilton, Ontario, L8S 4M1, Canada}
\author{J. Gaudet}
\affiliation{Department of Physics and Astronomy, McMaster University, Hamilton, Ontario, L8S 4M1, Canada}
\author{K. Fritsch}
\affiliation{Department of Physics and Astronomy, McMaster University, Hamilton, Ontario, L8S 4M1, Canada}
\affiliation{Helmholtz-Zentrum Berlin f\"ur  Materialien und Energie, Hahn-Meitner-Platz 1, 14109 Berlin, Germany}
\author{D.~Pomaranski}
\affiliation{Department of Physics and Astronomy and Guelph-Waterloo Physics Institute, University of Waterloo, Waterloo, Ontario, Canada N2L 3G1}
\affiliation{Institute for Quantum Computing, University of Waterloo, Waterloo, Ontario, Canada N2L 3G1}
\author{J. B. Kycia}
\affiliation{Department of Physics and Astronomy and Guelph-Waterloo Physics Institute, University of Waterloo, Waterloo, Ontario, Canada N2L 3G1}
\affiliation{Institute for Quantum Computing, University of Waterloo, Waterloo, Ontario, Canada N2L 3G1}
\author{Y. Qiu}
\affiliation{NIST Center for Neutron Research, NIST, Gaithersburg, Maryland 20899-8102, USA}
\author{J. R. D. Copley}
\affiliation{NIST Center for Neutron Research, NIST, Gaithersburg, Maryland 20899-8102, USA}
\author{M. Couchman}
\affiliation{Department of Physics and Astronomy, McMaster University, Hamilton, Ontario, L8S 4M1, Canada}
\author{A. Morningstar}
\affiliation{Department of Physics and Astronomy, McMaster University, Hamilton, Ontario, L8S 4M1, Canada}
\author{H. A. Dabkowska}
\affiliation{Brockhouse Institute for Materials Research, Hamilton, Ontario, L8S 4M1, Canada}
\author{B. D. Gaulin}
\affiliation{Department of Physics and Astronomy, McMaster University, Hamilton, Ontario, L8S 4M1, Canada}
\affiliation{Brockhouse Institute for Materials Research, Hamilton, Ontario, L8S 4M1, Canada}
\affiliation{Canadian Institute for Advanced Research, 180 Dundas St.\ W., Toronto, Ontario, M5G 1Z8, Canada}

\date{\today}

\begin{abstract}
Recent low temperature heat capacity (C$_P$) measurements on polycrystalline samples of the pyrochlore antiferromagnet Tb$_{2+x}$Ti$_{2-x}$O$_{7+\delta}$ have shown a strong sensitivity to the precise Tb concentration $x$, with a large anomaly exhibited for $x \sim 0.005$ at $T_C \sim 0.5$ K and no such anomaly and corresponding phase transition for $x \le 0$.  We have grown single crystal samples of Tb$_{2+x}$Ti$_{2-x}$O$_{7+\delta}$, with approximate composition $x=-0.001, +0.0042$, and $+0.0147$, where the $x=0.0042$ single crystal exhibits a large C$_P$ anomaly at $T_C$=0.45 K, but neither the $x=-0.001$ nor the $x=+0.0147$ single crystals display any such anomaly.  We present new time-of-flight neutron scattering measurements on the $x=-0.001$ and the $x=+0.0147$ samples which show strong $\left(\frac{1}{2},\frac{1}{2},\frac{1}{2}\right)$ quasi-Bragg peaks at low temperatures characteristic of short range antiferromagnetic spin ice (AFSI) order at zero magnetic field but only under field-cooled conditions, as was previously observed in our $x = 0.0042$ single crystal.  Furthermore, the frozen AFSI state displays a gapped spin excitation spectrum around $\left(\frac{1}{2},\frac{1}{2},\frac{1}{2}\right)$, with a gap of $\sim$ 0.1 meV, again similar to previous observations on the $x = 0.0042$ single crystal.  These results show that the strong $\left(\frac{1}{2},\frac{1}{2},\frac{1}{2}\right)$ quasi-Bragg peaks and gapped AFSI state at low temperatures under field cooled conditions are robust features of Tb$_2$Ti$_2$O$_7$, and are not correlated with the presence or absence of the C$_P$ anomaly and phase transition at low temperatures.  Further, these results show that the ordered state giving rise 
to the C$_P$ anomaly is confined to $0 \leq x \leq 0.01$ for  Tb$_{2+x}$Ti$_{2-x}$O$_{7+\delta}$, 
and is not obviously connected with conventional order of magnetic dipole degrees of freedom.
\end{abstract}


\maketitle

\section{Introduction}
%
Geometric frustration is a natural route towards the stabilization of exotic quantum states in magnetic materials.\cite{IFM}  Considerable effort has been directed to the study of the cubic pyrochlore lattice, one of the canonical architectures supporting geometrical frustration in three dimensions.\cite{Gardner2010}  Such materials have the chemical composition A$_2$B$_2$O$_7$ with both the A$^{3+}$ sublattice and the B$^{4+}$ sublattice, independently, occupying the vertices of corner-sharing tetrahedra. The rare earth titanate family $R_2$Ti$_2$O$_7$ ($R=$rare-earth) remains a source of surprise and excitement even after more than a decade of extensive study.  The Ti$^{4+}$ ions are non-magnetic within this structure and the magnetism originates from trivalent rare-earth ions on the A pyrochlore sublattice.  These magnetic moments can display varied interactions and anisotropies resulting in very rich and exotic phenomena such as emergent magnetic monopole excitations and a range of interesting phase behavior at low temperatures displaying spin ice\cite{Bramwell2001,Morris2009} and spin liquid ground states\cite{Hermele2004,Savary2012QSL,Benton2015}, as well as order selected by order-by-disorder mechanisms.\cite{Savary2012OBD}

The pyrochlore antiferromagnet Tb$_2$Ti$_2$O$_7$ has long been a promising quantum spin liquid candidate as it fails to find a conventional magnetically-ordered state down to at least 20 mK.\cite{Gardner99} However, the exact nature of its disordered ground-state remains of some debate \cite{Fennell2012,Fennell2014,Guitteny2013}. It displays an antiferromagnetic Curie-Weiss susceptibility with a $\Theta_{CW} \sim$ -17 K, and the crystalline electric field $J=6$ eigenstates of the Tb$^{3+}$ ion give rise to a ground state and a first excited state doublets that are separated by only $\sim$ 1.5 meV.  While the corresponding eigenfunctions correspond primarily to m$_J$=$\pm$4 and $\pm$ 5,\cite{Zhang2014,Princep2015,Mirebeau2007} and therefore to Ising-like anisotropy, the proximity in energy of the ground and first excited state doublets makes it difficult to consider a $S_{\rm eff}=\frac{1}{2}$ description of the moments, and therefore precludes the relatively simple spin Hamiltonian which can result from a well-isolated ground state doublet.  

At very low temperatures, Tb$_2$Ti$_2$O$_7$ is known to display a peak in the real ac susceptibility between $T_g$ $\sim$ 0.15 K and 0.25 K depending on both sample and frequency, and this is consistent with a glassy ground state below $T_g$. \cite{Yin2013,Hamaguchi2004,Lhotel2012}  Anomalies in the heat capacity, $C_P$, indicative of a thermodynamic phase transition, are also observed at higher critical temperatures, $T_C$ between 0.4 K and 0.5 K in a subset of samples.  Our own earlier neutron scattering work on a single crystal of Tb$_2$Ti$_2$O$_7$ with a large C$_P$ anomaly at $T_C =0.45$~K (labelled from now on as sample B), was observed to display quasi-Bragg peaks at ordering wavevectors of the form $\left(\frac{1}{2},\frac{1}{2},\frac{1}{2}\right)$ below $T_g \sim 0.275$~K, with a concomitant gap in its spin excitation spectrum of $\sim$ 0.08 meV.\cite{Fritsch2013}  This state is stabilized at zero magnetic field only under a magnetic field-cooled protocol, and can be understood in terms of a short-ranged antiferromagnetic spin ice (AFSI) structure, wherein an ordered correlated region is comprised of $\sim$ 8 conventional unit cells which individually adopt ordered spin ice domains, with two spins pointing in and two spins pointing out of each tetrahedron.  These domains are then antiferromagnetically correlated between neighbouring unit cells. Another symmetry allowed spin structure made of consecutive layers of one-in/three-out, two-in/two-out and three-in/one-out tetrahedra was recently found to share the same powder-averaged structure factor. However, the elastic diffuse scattering measured on single crystals suggests that the AFSI structure is stabilized at low temperature.\cite{Guitteny2015}

Recent work on polycrystalline samples of Tb$_{2+x}$Ti$_{2-x}$O$_{7+\delta}$, with the off-stoichiometry $x$ tightly controlled,  has demonstrated a strong sensitivity of $T_C$ to the precise Tb concentration.\cite{Taniguchi2013}  In particular this work showed that the size of the C$_P$ anomaly is maximal at $x \sim 0.005$ and $T_C$ $\sim$ 0.5 K and that no such C$_P$ anomaly and corresponding phase transition is observed at any temperature for $x \le 0$.  We were then motivated to investigate the $x$-dependence of the AFSI ground state and its gapped spin excitation spectrum using single crystals, and that is what we report here.  To this end, we grew two new single crystal samples, hereafter referred to as samples A and C, which correspond to Tb$_{2+x}$Ti$_{2-x}$O$_{7+\delta}$ with $x=-0.0010$ and $x= +0.0147$.  In this paper we examine the systematics of the elastic and inelastic magnetic neutron scattering in these two new crystals as compared with our previous results for sample B, with $x=+0.0042$.
 
\begin{figure}
\includegraphics[width=\columnwidth]{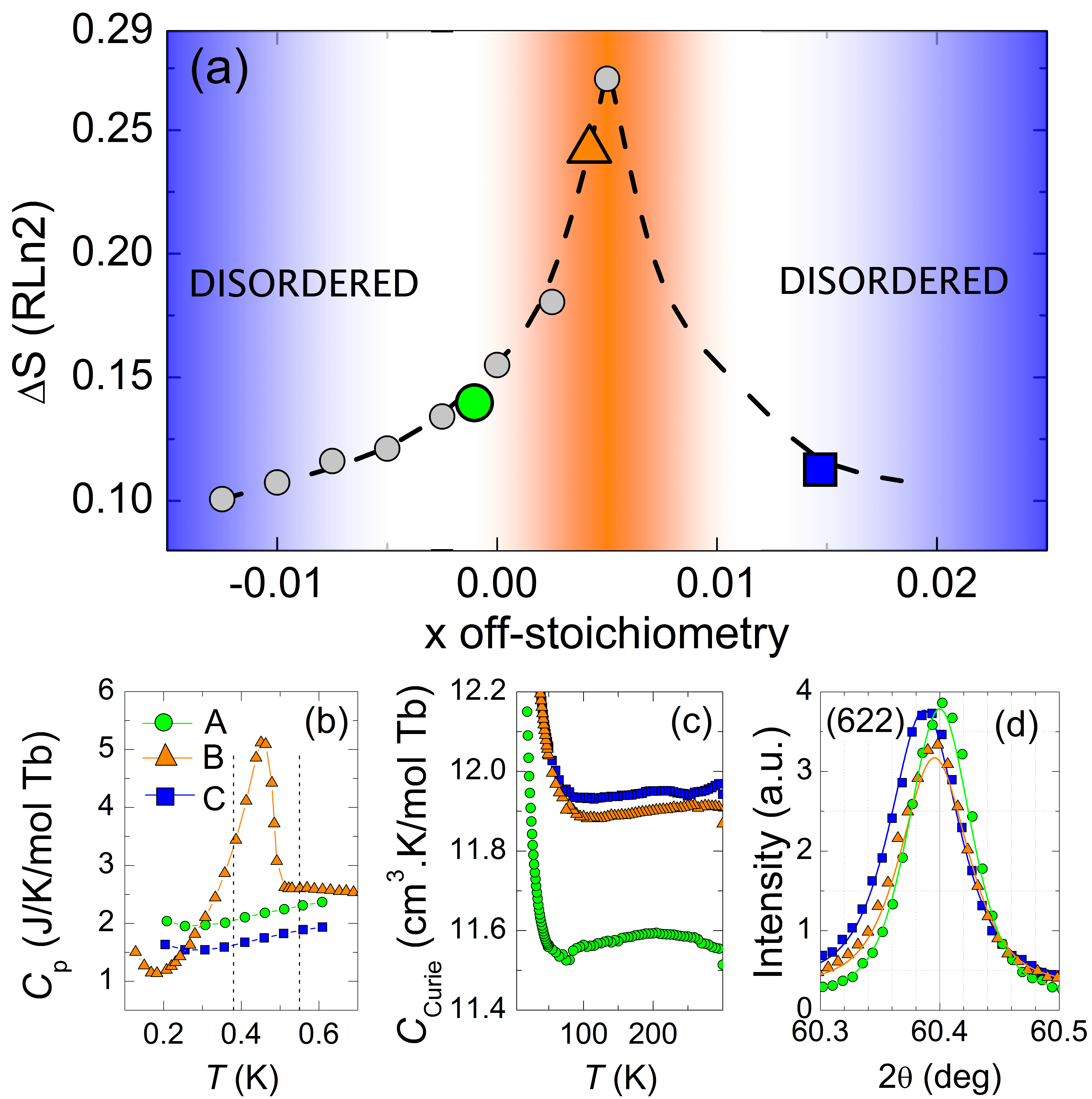}
\caption{\label{phasediag} (Color online)
(a) The phase diagram for Tb$_{2+x}$Ti$_{2-x}$O$_{7+\delta}$ showing the release of entropy between 0.38 and 0.55~K as a function of the off-stoichiometry $x$.  The grey disks are from measurements on polycrystalline samples, from Ref.\onlinecite{Taniguchi2013}. Dashed line is a guide to the eye.
(b) Temperature dependence of the specific heat ($C_P$) for samples \textit{A}, \textit{B} and \textit{C}. The dashed lines delimit the integration range used for the entropy calculation.
(c) Temperature dependence of the Curie constant ($C_{\rm Curie}$) for samples \textit{A}, \textit{B} and \textit{C}. 
(d) High resolution x-ray powder diffraction data showing the region around the (622) Bragg peak for samples A, B and C.
}
\end{figure}
%

\section{Experimental Details }

\begin{table*}
\caption{\label{Tablesum} Summary of the characterization of the three single crystal samples \textit{A}, \textit{B} and \textit{C}, showing the Curie constant of the magnetic susceptibility ($C_{\rm Curie}$), the lattice constant $a$, the value of the off-stoichiometry $x$ in the chemical formula Tb$_{2+x}$Ti$_{2-x}$O$_{7+\delta}$, the ratio of the elastic neutron scattering intensity of the $(\frac{1}{2},\frac{1}{2},\frac{5}{2})$ magnetic quasi-Bragg peak to that of the $(113)$ nuclear Bragg peak, the moment related to the AFSI order and the correlation length.}
\begin{ruledtabular}
\begin{tabular}{cccccccc}
Sample      & $C_{\rm Curie}$(cm$^3$.K/mol Tb) & $a$({\AA}) & $x$ & Int.$(\frac{1}{2},\frac{1}{2},\frac{5}{2})$/$(113)$ & moment($\mu_B$)  & $\xi$({\AA})& \\
\hline
\textit{A}  & 11.58(2)   & 10.15784(2) & $-0.0010(2)$ & 0.32 &  1.8(2) & 28(2) &  \\
\textit{B}  & 11.90(2)   & 10.15849(2) & $+0.0042(2)$ & 0.66 &  2.5(3) & 28(2) & \\
\textit{C}  & 11.95(2)   & 10.15980(2) & $+0.0147(2)$ & 0.22  &   1.5(2) & 21(4) &  \\
\end{tabular}
\end{ruledtabular}
\end{table*}

Two single-crystal samples of Tb$_2$Ti$_2$O$_7$ were grown using the optical floating zone technique at McMaster University\cite{Dabkowska2010,Gardner1998}. These crystals, which are referred to as samples A and C, were intentionally grown off-stoichiometry; with target compositions of Tb$_{2+x}$Ti$_{2-x}$O$_{7+\delta}$ with $x=-0.005$ and $+0.005$, under an absolute pressure of 3 atm of O$_2$. We compare our results on these single crystal samples with sample B, which was previously grown by the same optical floating zone technique at McMaster University, but with a target stoichiometric $x=0$ composition.  We carried out characterization studies with the intent of accurately estimating $x$ for each of the three samples.  

High resolution x-ray scattering measurements were carried out at McMaster University, using a PANalytical powder diffractometer equipped with a X'Celerator detector\cite{NISTDisclaimer}. Cu-K$_{\alpha1}$ x-rays ($\lambda = 1.540598$ \AA) were generated from a Cu anode.  For each of the three single crystals, small discs were cut from the samples used for neutron scattering measurements and then ground in an agate pestle and mortar into a $\sim 100$ mg uniform powder. A small amount of pure silicon powder ($\sim 10$ mg) was added to serve as a common reference for the determination of the lattice parameters. Diffraction patterns were measured at room temperature with the 2$\theta$ range extending from 10 to 120 degrees, with a step size of 0.00835 degrees, and then refined using the FULLPROF package.\cite{Fullprof} 

New specific heat measurements were performed on samples A and C with a $^3$He/$^4$He dilution refrigerator at the University of Waterloo. The single crystals used for neutron scattering measurements were cut and sectioned into $\sim$ 30 mg masses with dimensions 2 $\times$ 2 $\times$ 2 mm$^3$ and these smaller single crystals were used in the heat capacity study. The relaxation method was employed with a thermal weak link of manganin wire, with conductance 5.0 $\times$ 10$^{-7}$ J/K/s at 0.80 K. The resulting time constant was greater than 600 s at the highest temperature measured. The average step size for the relaxation measurement was 3.5\% of the nominal temperature, with a minimum equilibration time window of five times the thermal relaxation constant.  These results are combined here with our earlier measurements on sample B, which employed very similar protocols.

Magnetic susceptibility measurements were performed on $\sim$ 4 mg  and 100 mg powder samples taken from each of single crystals A, B and C, and using an applied magnetic field of $\mu_0 H = 1$~T.  The measurements employed a Quantum Design SQUID magnetometer operating between 1.8 K and 300 K\cite{NISTDisclaimer}.

Time-of-flight neutron scattering measurements were performed using the Disk Chopper Spectrometer (DCS)  at the NIST Center for Neutron Research.\cite{Copley2003} Two different incident energies were employed; for lower energy resolution measurements we employed $E_i = 3.27$~meV, giving an energy resolution of 0.1 meV, while the higher energy resolution measurements employed $E_i = 1.28$ meV with a resulting resolution of 0.02 meV. The sample was carefully aligned with the $[1\overline{1}0]$ direction vertical to within 0.5 degrees, and mounted in a dilution refrigerator magnet cryostat such that the (H,H,L) plane was coincident with the horizontal scattering plane. Measurements were performed in a temperature range between 80 mK and 650 mK, and employed magnetic fields of up to 0.2 T applied along the vertical $[1\overline{1}0]$ direction.

\section{Results and Discussion}
\subsection{Bulk Characterization and Structure }

Heat capacity measurements carried out on single crystal samples A, B, and C of Tb$_{2+x}$Ti$_{2-x}$O$_{7+\delta}$ are shown in Fig. \ref{phasediag}b) for temperatures between 0.2 K and 0.6 K for samples A and C, and between 0.12 K and 0.7 K for sample B.  As mentioned above, sample B is the single crystal sample that was grown with starting constituents aimed at producing a stoichiometric, $x=0$, sample.  Both neutron scattering and heat capacity measurements have already been reported for it \cite{Fritsch2014}, and its $C_P$ data is reproduced in Fig. \ref{phasediag}b) for ease of comparison.  Clearly it shows a large $C_P$ anomaly at $T_C =0.45$ K, indicative of a thermodynamic phase transition.  Samples A and C are new single crystals that were grown from starting constituents aimed at producing off-stoichiometry crystals, $x=-0.005$ and $+0.005$, respectively.  As can be seen in Fig. \ref{phasediag}b), neither of the samples A and C displays a C$_P$ anomaly within this field of view.

In Fig.\ref{phasediag}a) we compare the integrated heat capacity, or release of entropy, in our three single crystals, over the temperature range from $T=0.38$ K to 0.55 K.  This temperature range covers the range of the C$_P$ anomalies previously reported in polycrystalline samples.  The integrated C$_P$ from our single crystal samples is also compared with C$_P$ integrated over the same temperature range from a set of powder samples from Taniguchi \textit{et al.}, with stoichiometries estimated to range from -0.013 to +0.005 (grey disks in Fig. \ref{phasediag}a)).\cite{Taniguchi2013,Taniguchi2015} 
This comparison allows us to estimate our actual single crystal stoichiometries, as opposed to those expected on the basis of the masses of the starting materials in the single crystal growth, relative to those determined by Taniguchi \textit{et al.}
%
From this comparison, we estimate the stoichiometry for single crystal sample B as  $x= +0.0042(2)$. The stoichiometries of samples A and C are then inferred from the values of their lattice parameters determined by x-ray diffraction measurements (see below), and are found to be $x=-0.0010(2)$ and +0.0147(2), respectively.
We also note that only one of our single crystal samples, sample B with $x=0.0042$, displays a $C_P$ anomaly.  In Fig. 1 a) we therefore label the approximate range in stoichiometry from $x=0$ to $+0.01$ as corresponding to an ordered ground state with some sort of thermodynamic phase transition, while those stoichiometries outside of this narrow bound display the corresponding disordered ground states, at least for temperatures above $T=0.2$~K.

Figure \ref{phasediag}c) shows the temperature dependence of the Curie constant extracted from a Curie-Weiss fit of the magnetic susceptibility $\chi = \chi_0 + C_{\rm Curie}/(T - \theta_{\rm CW})$, measured in a $\mu_0 H=1$~T magnetic field for crushed powders of each of our single crystal samples. The temperature independent term $\chi_0$ comprises the core diamagnetic and the Van Vleck paramagnetic contributions. Its common value of $1.5(1).10^{-3}$~cm$^3$/mol Tb is shared between the three data sets. The Curie-Weiss temperatures $\theta_{\rm CW}$ vary between $-12$~K and $-14$~K, while the Curie constants themselves, extracted from the fits, are 11.58(2), 11.90(2) and 11.95(2) in units of cm$^{3}$.K/mol Tb (Table \ref{Tablesum}). It is clear that the Curie constant increases from sample A to B to C, reflecting an increase in the off-stoichiometry $x$, and hence the Tb-content.

Figure \ref{phasediag}d) shows high resolution x-ray powder diffraction data taken on small pieces of single crystal samples A, B, and C, which were subsequently crushed to form a powder and then mixed with a small amount of pure Si powder.  The room temperature powder x-ray diffraction measurements were taken with Cu-K$_{\alpha1}$ radiation, and inclusion of the Si powder, with a very accurately known lattice parameter, allows an accurate room temperature measurement of the cubic lattice parameters for our single crystal Tb$_{2+x}$Ti$_{2-x}$O$_{7+\delta}$ samples, on an absolute scale.  Figure \ref{phasediag}d) highlights the powder diffraction pattern around the (662) Bragg peak of Tb$_2$Ti$_2$O$_7$. No shift is observed in the reference Bragg signal from the Si powder, and the Rietveld refinements of the entire diffraction patterns give the room temperature (293 K) lattice parameters of our A, B and C single crystals of Tb$_{2+x}$Ti$_{2-x}$O$_{7+\delta}$ as 10.15784(2)~\AA, 10.15849(2)~\AA~ and 10.15980(2)~\AA, respectively. While these values are slightly larger than those determined for the powder samples prepared by Taniguchi \textit{et al.}, the relative change in lattice parameter with stoichiometry $x$ is approximately the same. Assuming that the slope of the linear relationship between the cubic lattice parameter and the off-stoichiometry $x$ reported in Ref.\onlinecite{Taniguchi2015} holds, the compositions of our samples A, B and C are respectively $x=-0.0010(2), +0.0042(2)$ and +0.0147(2) (Table \ref{Tablesum}).

\subsection{Neutron Scattering}

\begin{figure}
\includegraphics[width=\columnwidth]{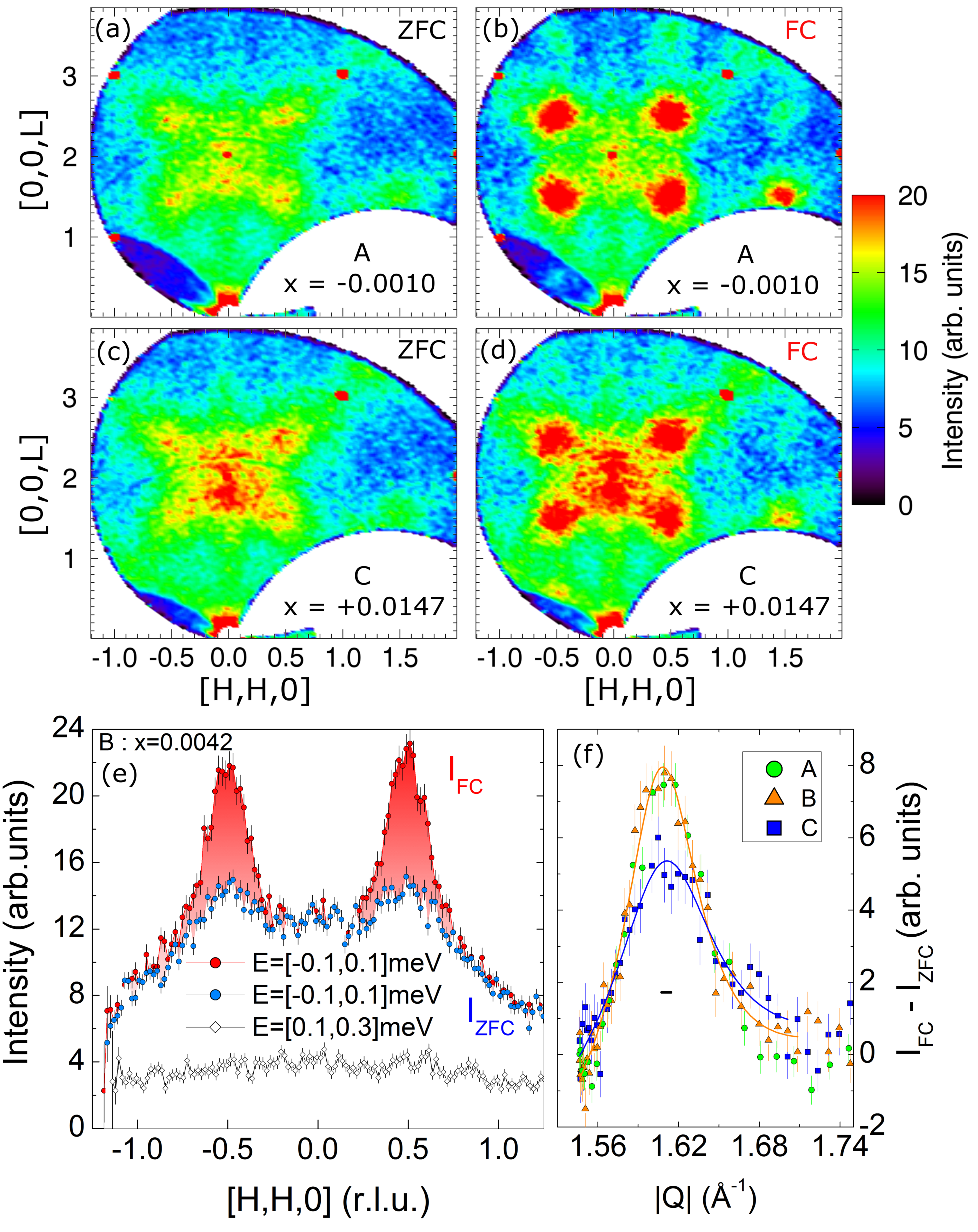}
\caption{\label{Fig_Emaps} (Color online) Panels a) and b) show reciprocal space maps of elastic neutron scattering ($\pm$ 0.1 meV) within the HHL plane for sample A, Tb$_{2+x}$Ti$_{2-x}$O$_{7+\delta}$ with $x=-0.0010(2)$ taken at $T=0.1$~K under ZFC (a) and FC (b) protocols.  Panels c) and d) show similar data for sample C, $x=+0.0147(2)$.
%
(e) {\bf Q}-scans through the elastic neutron data shown in panels a) and b) of the form HHL, where L is integrated from 2.3 to 2.8, for sample B at $T=0.1$~K.  Elastic scattering ($\pm$ 0.1 meV) scans are shown under both FC and ZFC protocols.  A similar {\bf Q}-scan of the inelastic scattering (0.1 meV $\le E \le 0.3$~meV) is shown for comparison.  Panel f) shows the difference between the FC and ZFC {\bf Q} scans at $T=0.1$~K for all three single crystals of Tb$_{2+x}$Ti$_{2-x}$O$_{7+\delta}$, A, B, and C.  Lorentzian fits to the resulting lineshape, from which correlation lengths can be estimated, are shown, along with a horizontal bar representing the resolution limit of such a measurement. Error bars in all figures represent one standard deviation.}
\end{figure}

\begin{figure}
\includegraphics[width=\columnwidth]{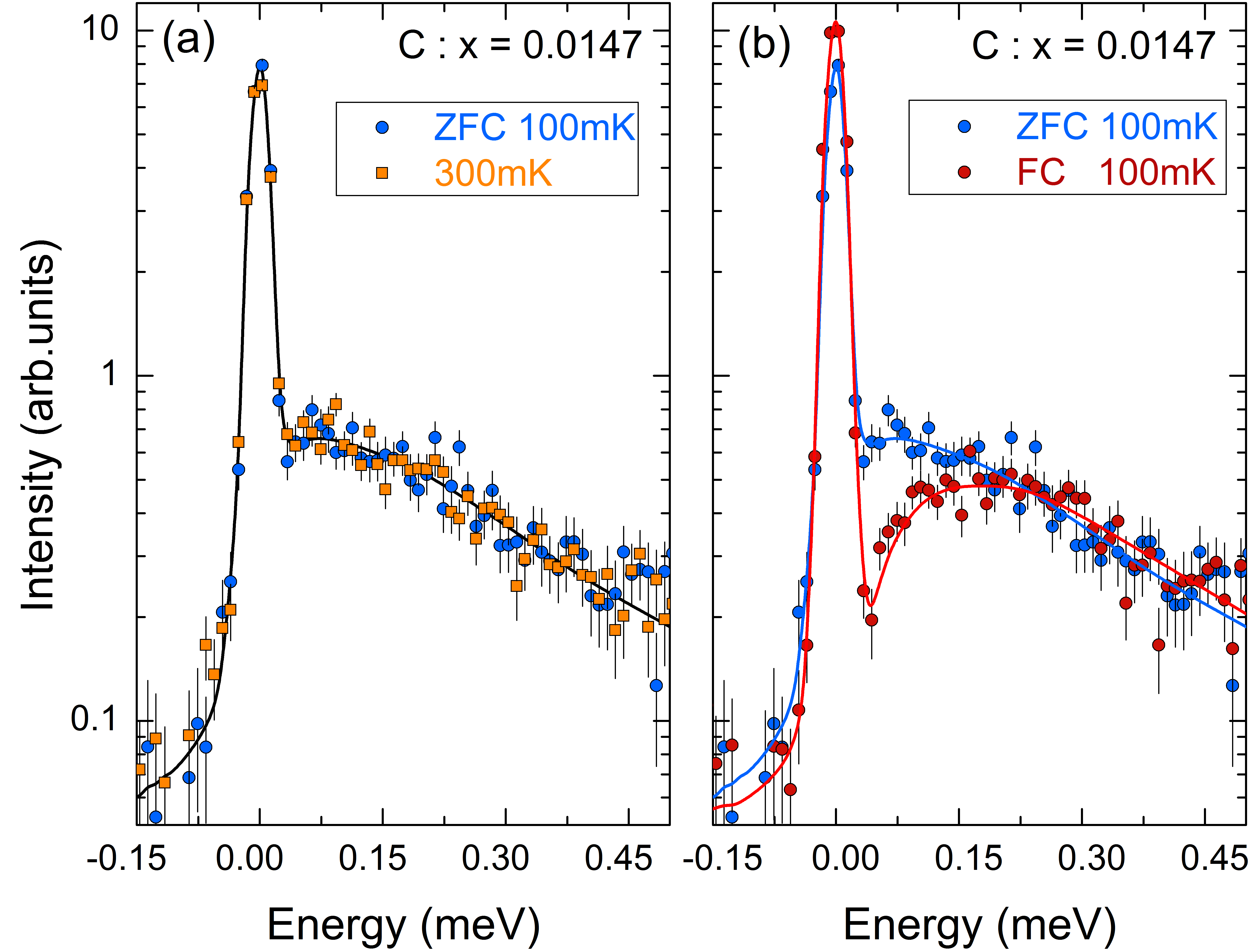}
\caption{\label{Fig_Ecuts} (Color online) High resolution energy scans of an integration around the $(\frac{1}{2},\frac{1}{2},\frac{3}{2})$ quasi-Bragg peak in sample C, which displays no low temperature C$_P$ anomaly.  Panel a) shows data taken at $T=0.1$~K with a ZFC protocol, and at $T=0.3$~K.  Both data sets show identical elastic scattering, and display a gapless quasi-elastic inelastic spectrum.  Panel b) shows data taken at $T=0.1$~K under both a FC and a ZFC protocol.  One clearly sees the increased elastic scattering giving rise to the $(\frac{1}{2},\frac{1}{2},\frac{3}{2})$ quasi-Bragg peak under FC conditions, along with the concomitant gapped spin excitation spectrum with a gap of $\sim$ 0.1 meV. Error bars in both panels represent one standard deviation.}
\end{figure}

New time-of-flight neutron scattering data on single crystal Tb$_{2+x}$Ti$_{2-x}$O$_{7+\delta}$ samples A and C are shown in Figs. \ref{Fig_Emaps} and \ref{Fig_Ecuts}, where Fig. \ref{Fig_Emaps} shows elastic scattering data with an energy resolution of 0.1 meV, and Fig. \ref{Fig_Ecuts} shows both elastic and inelastic neutron scattering with an energy resolution of 0.02 meV.  Figure \ref{Fig_Emaps} shows reciprocal space maps of the elastic scattering within the (HHL) plane for both samples A and C, under zero field cooled (ZFC) and field cooled (FC).  These data sets show intense magnetic quasi-Bragg peaks to appear at $(\frac{1}{2},\frac{1}{2},\frac{1}{2})$ and symmetry related positions for both samples A and C, under FC conditions only.  For these FC data sets, the samples were cooled from $T=4$~K in a $\mu_0 H=0.2$~T field aligned along the $[1\overline{1}0]$ direction, while the ZFC data sets were obtained while cooling in nominal zero field, where only the residual magnetic field of the superconducting magnet cryostat (typically 0.01 T) is present.  The quasi-Bragg peaks are seen to form below $T_{\rm AFSI}$ $\sim$ 275 mK, consistent with earlier measurements on sample B.  The low temperature elastic reciprocal space maps obtained under ZFC conditions resemble the FC maps, although we observe only vestiges of the quasi-Bragg peaks at $(\frac{1}{2},\frac{1}{2},\frac{1}{2})$ positions.  These features are much more diffuse than those observed under FC conditions, and a ``checkerboard'' pattern of ZFC diffuse scattering is observed such that $(\frac{1}{2},\frac{1}{2},\frac{1}{2})$ reciprocal space positions occupy the corners of the diffuse scattering ZFC checkerboard.  This is again consistent with what was observed earlier on sample B of Tb$_{2+x}$Ti$_{2-x}$O$_{7+\delta}$ under ZFC conditions, and from other studies which did not employ FC protocols\cite{Petit2012, Guitteny2013}.

Measurements using different FC cooling protocols (in $[1\overline{1}0]$] field from high temperatures; ZFC to low temperatures, then apply a field and remove it), demonstrate that it is only important to have applied a magnetic field along $[1\overline{1}0]$ $\mu_0 H \geqslant 0.2$ T while the sample is at low temperatures, below  $T_{\rm AFSI}$ $\sim$ 275 mK, in order to observe intense quasi-Bragg peaks at $(\frac{1}{2},\frac{1}{2},\frac{1}{2})$ positions.

With the elastic neutron scattering data in hand, we can take cuts of the diffuse scattering for the purpose of extracting correlation lengths of the FC AFSI state in all three of samples A, B, and C.  Typical reciprocal space cuts of this data for sample B are shown in Fig. \ref{Fig_Emaps}e), where we show diffuse elastic scattering along the [H,H,0] direction, with an integration along L from L=2.3 to L=2.8, which is sufficient to cover the reciprocal space extent of the quasi-Bragg peaks at $(-\frac{1}{2},-\frac{1}{2},\frac{5}{2})$ and $(\frac{1}{2},\frac{1}{2},\frac{5}{2})$.  
For comparison we have overplotted low energy inelastic scattering, integrating from 0.1 meV to 0.3 meV under both FC and ZFC conditions (averaged together).  Consistent with the reciprocal space maps in Fig. \ref{Fig_Emaps}, we see much sharper quasi-Bragg peaks present at low temperature under FC conditions only.

Figure \ref{Fig_Emaps}f) shows the difference between FC and ZFC elastic scattering along [HH0], as defined in Fig. \ref{Fig_Emaps}e), and from this we can extract the low temperature correlation length induced by FC protocols relative to ZFC protocols.  This difference scattering profile is shown for all three Tb$_{2+x}$Ti$_{2-x}$O$_{7+\delta}$ crystals, and we obtain somewhat longer correlation lengths, $\xi=28(2)$~\AA ~for samples A and B, than for sample C where we obtain $\xi = 21(4)$~\AA.  For ease of comparison, we have listed these correlation lengths for the three samples in Table \ref{Tablesum}.  These correlation lengths are larger than those previously estimated for sample B, which was $\xi \sim 8$~\AA.  However that correlation length corresponded to the inverse width in reciprocal space of the full diffuse elastic scattering around $(\frac{1}{2},\frac{1}{2},\frac{1}{2})$ type positions, as opposed to the inverse width of the difference between the FC and ZFC elastic diffuse scattering that we are presenting in Fig. \ref{Fig_Emaps}f).  Nonetheless we conclude that the local AFSI structure at low temperatures under FC conditions is a robust characteristic of Tb$_2$Ti$_2$O$_7$, and the local correlation volume corresponds to a small number ($\sim$ 2) of unit cells of the local AFSI structure, independent of the precise stoichiometry.

It is also possible to estimate the magnetic moment participating in the $(\frac{1}{2},\frac{1}{2},\frac{1}{2})$ quasi-Bragg peaks and to examine how that varies with off-stoichiometry, $x$.  To do this we integrate over the magnetic $(\frac{1}{2},\frac{1}{2},\frac{5}{2})$ quasi-Bragg peak obtained under FC protocols, in a transverse scan which covers the full extent of the quasi-Bragg peak, and we do the same thing for the nuclear-allowed (113) Bragg peak.  We can then calculate the static magnetic moment involved, assuming the AFSI local structure and the cubic pyrochlore structure. Previous measurements on powder samples by Taniguchi \textit{et al.} reported the existence of a true (resolution-limited) long range order for their $x=+0.005$ sample that showed a C$_P$ anomaly at $\sim$ 0.5 K, involving a very small static moment of 0.08~$\mu_B$/Tb\cite{Taniguchi2013}. With our new measurements it is qualitatively clear that a much larger static moment is present in a short range AFSI structure for all three single crystals of Tb$_{2+x}$Ti$_{2-x}$O$_{7+\delta}$,  as the $(\frac{1}{2},\frac{1}{2},\frac{1}{2})$ quasi-Bragg peaks are strong and easy to observe in all three single crystals.  Structure factor calculations comparing the integrated intensity of the $(\frac{1}{2},\frac{1}{2},\frac{5}{2})$ quasi-Bragg peak and the (113) nuclear Bragg peak produce static Tb moments of 1.8(2) $\mu_B$, 2.5(3) $\mu_B$, and 1.5(2) $\mu_B$, for single crystals A, B, and C of Tb$_{2+x}$Ti$_{2-x}$O$_{7+\delta}$ with $x=-0.0010(2)$, $+0.0042(2)$ and $+0.0147(2)$, respectively.  These estimates for the static moment in the quasi-Bragg peak are listed in Table \ref{Tablesum}.

Previous measurements on sample B, Tb$_{2+x}$Ti$_{2-x}$O$_{7+\delta}$ with $x=+0.0042$, showed that the appearance of the $(\frac{1}{2},\frac{1}{2},\frac{1}{2})$ quasi-Bragg peaks under FC protocols is coincident with the opening of a $\sim 0.08$~meV gap in the spin excitation spectrum.  As our new measurements on samples A and C shown in Fig.\ref{Fig_Ecuts} demonstrate, a very similar phenomenon is relevant to other stoichiometries as well, both with and without a low temperature C$_P$ anomaly.  Figure \ref{Fig_Ecuts} shows high energy-resolution inelastic neutron scattering, integrating around the $(\frac{1}{2},\frac{1}{2},\frac{3}{2})$ quasi-Bragg peak in sample C, Tb$_{2+x}$Ti$_{2-x}$O$_{7+\delta}$ with $x=+0.0147(2)$.  Figure \ref{Fig_Ecuts} a) shows inelastic data taken under ZFC protocols at $T=0.1$~K as well as $T=0.3$~K.  One sees that both the elastic and inelastic spectra are identical, and the inelastic magnetic scattering shows gapless, quasi-elastic scattering, down to our resolution limit, $\sim 0.02$~meV.   In contrast, data sets on sample C at $T=0.1$ K taken under FC and ZFC protocols show dramatic differences.  Larger elastic magnetic scattering is clearly seen near zero energy, and the spin excitation spectrum is gapped below $\sim 0.1$~meV.  Both of these results are consistent with earlier results on sample B \cite{Fritsch2013,Fritsch2014}.  Thus the appearance of $(\frac{1}{2},\frac{1}{2},\frac{1}{2})$ quasi-Bragg peaks and a concomitant gapping of the spin excitation spectrum under FC protocols for temperatures less than $\sim$ 0.275 K, are robust features of Tb$_2$Ti$_2$O$_7$, and are not correlated with the observation (or lack thereof) of C$_P$ anomalies at $\sim$ 0.4 K to 0.5 K.   

These results reaffirm our original assertion that the thermodynamic phase transition characterized in Tb$_{2+x}$Ti$_{2-x}$O$_{7+\delta}$ with $0 \leq x \leq 0.01$ by a C$_P$ anomaly near $T_C=0.4$ K to 0.5 K, is different from and independent of the transition to a short ranged AFSI state below $T_{\rm AFSI} \sim$ 0.275 K.  Our original assertion was made on the basis of the difference between $T_C$ and $T_{\rm AFSI}$.  The current work explicitly shows the $(\frac{1}{2},\frac{1}{2},\frac{1}{2})$ quasi-Bragg peaks and concomitant gap in the spin excitation spectrum occur in the absence of $T_C$.  As neutron scattering sees magnetic dipole degrees of freedom only, this implies that $T_C$ corresponds to degrees of freedom which are of higher order than dipole.  

This then begs the following question: why is $T_C$ and the ordered state connected with such quadrupoles or octupoles, \textit{etc}., so sensitive to the weak disorder connected with stoichiometry on the 0.5\% scale or less ?  The observation that the spin dipole degrees of freedom are robust to such off-stoichiometry is typical of most magnetic materials.  Higher-than-dipole order is likely associated with aspherical $f-$electron charge distributions, and that could be sensitive to the nature of the $4f$ electrons on the ``stuffed" B-site in pyrochlores of the form A$_2$B$_2$O$_7$.  Such stuffed, B-site, Tb ions would experience different crystal field effects from the A-site Tb ions, and these may act as impurities with a different aspherical $f-$electron charge distribution that are effective in frustrating higher-than-dipole order.  Such a scenario of very different crystal field eigenfunctions for rare earth ions on A and B sublattices has recently been discussed for Yb$_{2+x}$Ti$_{2-x}$O$_{7+\delta}$\cite{Gaudet2015}.  Additionally, there is the question of why the ordered states of Tb$_{2+x}$Ti$_{2-x}$O$_{7+\delta}$ are not centered at $x=0$; indeed $T_C$ appears to be maximal at $x \sim +0.005$.  This would suggest that excess Ti is very detrimental to such a higher-than dipole ordered state.

\section{Conclusions}

We have performed new heat capacity, susceptibility and neutron scattering measurements on two single crystal samples (samples A and C), which we characterize as Tb$_{2+x}$Ti$_{2-x}$O$_{7+\delta}$ with $x=-0.0010(2)$ and $+0.0147(2)$.  Neither of these new single crystals displays a peak in C$_P$ above $\sim$ 0.2 K.  Neutron scattering measurements on these crystals show the same $(\frac{1}{2},\frac{1}{2},\frac{1}{2})$ quasi-Bragg peaks and concomitant 0.1 meV gap in the spin excitation spectrum to appear below $T_{\rm AFSI} \sim$ 0.275 K when cooled using a FC protocol. This phenomenology is very similar to that previously observed in sample B, which is characterized by Tb$_{2+x}$Ti$_{2-x}$O$_{7+\delta}$ with $x=+0.0042(2)$, and which does display a large C$_P$ anomaly at $T_C$=0.45 K.   These results show that the $(\frac{1}{2},\frac{1}{2},\frac{1}{2})$ quasi-Bragg peaks and associated AFSI local structure are robust features of Tb$_2$Ti$_2$O$_7$, and are not sensitive to precise stoichiometry.  In contrast, the ordered phase corresponding to $T_C$ is distinct from that associated with $T_{\rm AFSI}$, and is located in a small range of stoichiometry, near, but not coincident with perfect stoichiometry: it is found for Tb$_{2+x}$Ti$_{2-x}$O$_{7+\delta}$ with $0 \leq x \leq 0.01$. Very recent work carried out on single crystal samples reached similar conclusions regarding the existence of a finite region for the ordered phase in the $x - T$ phase diagram.\cite{Wakita2015}

These results are consistent with and substantially amplify our original suggestion that the thermodynamic phase transition below $T_C$ is associated with higher-than dipole order (quadrupole, octupole \textit{etc}.), while the AFSI state observed under FC protocols is associated with the magnetic dipoles.

\section*{Acknowledgments}
This work utilized facilities supported in part by the National Science Foundation under Agreement No. DMR-0944772, and was supported by NSERC of Canada. The DAVE software package\cite{DAVE} was used for data reduction and analysis of DCS data.


\begin{thebibliography}{99}
\bibitem{NISTDisclaimer}
The identification of  the commercial product does not indicate that it is endorsed by NIST or is recommended by NIST or that it is necessarily the best equipment for the purposes described.
\end{thebibliography}
\end{document}